\documentclass[aps,pra,reprint,amsmath,amssymb,superscriptaddress,longbibliography]{revtex4-2}
\usepackage{graphicx}
\usepackage[colorlinks=true,urlcolor=blue,citecolor=blue,linkcolor=blue,bookmarks=false, pdfstartview={FitH}]{hyperref}
\usepackage{amsmath}
\usepackage{bm}
\usepackage{braket}
\usepackage{dcolumn}
\usepackage{textcomp}
\usepackage{gensymb}
\usepackage{physics}
\usepackage{ragged2e}
\usepackage{color,soul}

\begin{document}

\title{Chiral electronic excitations in a quasi-two-dimensional Rashba system BiTeI}

\author{A.\,C.~Lee}
\email{aclee314@physics.rutgers.edu}
\affiliation{Department of Physics \& Astronomy, Rutgers University, Piscataway, New Jersey 08854, USA}
\author{B.~Peng}
\affiliation{Cavendish Laboratory, University of Cambridge, J. J. Thomson Avenue, Cambridge CB3 0HE, United Kingdom}
\author{K.~Du}
\affiliation{Department of Physics \& Astronomy, Rutgers University, Piscataway, New Jersey 08854, USA}
\affiliation{Rutgers Center for Emergent Materials, Rutgers University, Piscataway, New Jersey 08854, USA}
\author{H.-H.~Kung}
\affiliation{Department of Physics \& Astronomy, Rutgers University, Piscataway, New Jersey 08854, USA}
\author{B.~Monserrat}
\affiliation{Cavendish Laboratory, University of Cambridge, J. J. Thomson Avenue, Cambridge CB3 0HE, United Kingdom}
\affiliation{Department of Materials Science and Metallurgy, University of Cambridge, 27 Charles Babbage Road, Cambridge CB3 0FS, United Kingdom}
\author{S.\,W.~Cheong}
\affiliation{Department of Physics \& Astronomy, Rutgers University, Piscataway, New Jersey 08854, USA}
\affiliation{Rutgers Center for Emergent Materials, Rutgers University, Piscataway, New Jersey 08854, USA}
\affiliation{Laboratory for Pohang Emergent Materials and Max Planck POSTECH Center for Complex Phase Materials, Department of Physics, Pohang University of Science and Technology, Pohang 37673, Korea}
\author{C.\,J.~Won}
\affiliation{Laboratory for Pohang Emergent Materials and Max Planck POSTECH Center for Complex Phase Materials, Department of Physics, Pohang University of Science and Technology, Pohang 37673, Korea}
\author{G.~Blumberg}
\email{girsh@physics.rutgers.edu}
\affiliation{Department of Physics \& Astronomy, Rutgers University, Piscataway, New Jersey 08854, USA}
\affiliation{National Institute of Chemical Physics and Biophysics, 12618 Tallinn, Estonia}
\date{\today}

\begin{abstract}
The optical transitions between spin-polarized bands of the quasi-two-dimensional Rashba system BiTeI are investigated using polarization-resolved resonant Raman spectroscopy. 
We detect chiral excitations between states with opposite helicity and compare spectra to calculations within a three-band model. 
Using the resonant Raman excitation profile, we deduce the Rashba parameters and band gaps of the higher conduction bands near the Fermi level, and compare the parameters to values obtained by {\it ab initio} density functional theory. 
\end{abstract}

\maketitle 
The chirality of a physical state or an excitation is a configurational property of an object that cannot be superimposed on its mirror image, thereby imparting a handedness~\cite{cheong2018broken,Cheong2019,doi:10.1126/sciadv.abj8030}. 
In solids, the wave functions of chiral excitations are antisymmetric under mirror reflection. 
Examples include excitons in cuprates~\cite{liu1993novel,salamon1995largeshift,khveshchenko1994raman}, electron-hole excitations in graphene~\cite{Gallais_PRL2016}, collective spin modes in Dirac materials~\cite{Kung_PRL2017}, collective modes in the {\it hidden order} phase of exotic heavy fermion systems~\cite{Kung1339}, to name a few. 
The reciprocity breaking chiral excitations could result in three-spin excitations with a volume $S_i \cdot S_j \times S_k$~\cite{sulewski1991observation}, chiral spin modes~\cite{Kung_PRL2017}, circular charge current~\cite{shastry1990theory}, chiral excitons~\cite{liu1993novel,Kung2019PNAS}, or excitations of higher-order multipoles~\cite{Kung1339,SeanBook2022} which break mirror symmetries. 

Systems with strong spin-orbit coupling give rise to electronic bands with a helical spintexture, where the spin and momentum of the electron are intrinsically coupled. 
For systems where inversion symmetry is broken, the Rashba interaction~\cite{rashba1960ei} induces linear momentum-dependent splitting of the spin-degenerate electronic bands~\cite{PhysRevLett.92.126603,PhysRevB.91.035106,Bihlmayer_2015,PETERSEN200049,PhysRevLett.107.156803,PhysRevB.85.195401}. 
The resulting subbands are spin polarized with opposite helicity. 
The set of direct transitions between occupied and unoccupied Rashba-split bands of opposite helicity forms the excitation spectra of the electronic continuum of chiral symmetry, the ``Rashba continuum.'' 
We have chosen to investigate the quasi-two-dimensional (2D) Rashba system BiTeI using resonant electronic Raman scattering to gain deeper insight into the nature of the chiral Rashba continuum. 

\begin{figure}[t]%
	\centering
	\includegraphics[width=1.0\linewidth]{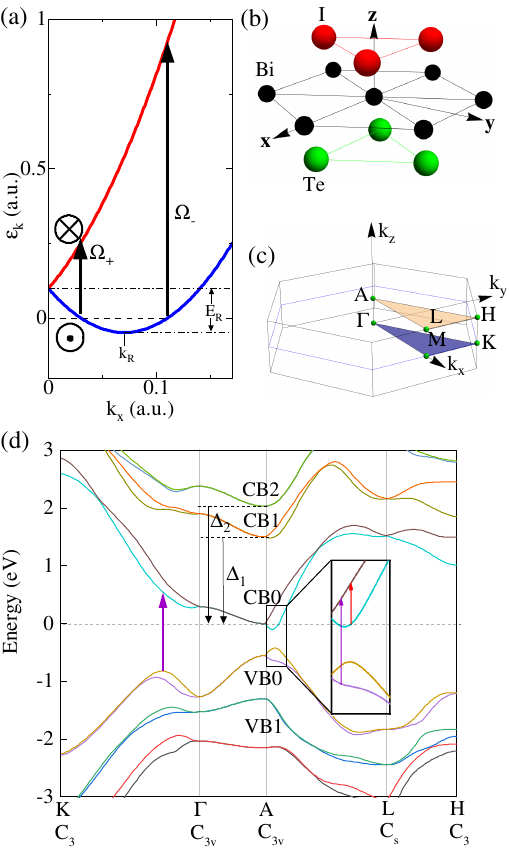}
	\caption{\textbf{Basic features of BiTeI}. 
	(a) The Rashba dispersion of a quasi-2D Rashba system. 
	The lower subband reaches a minimum at $k_{xy}=k_{R}$. 
	The energy difference between the $k_{xy}=0$ and $k_{R}$ points is $E_{R}$. 
	In this example, the spin orientation of the lower subband is directed out of page and the upper subband is directed in page. 
	The bounds of the Rashba continuum are indicated by arrows with the indices $\Omega_{+}$ and $\Omega_{-}$. 
	(b) The crystal structure of BiTeI. 
	(c) The Brillouin zone of the \textit{P3m1} space group. 
	(d) DFT calculation of the electronic band dispersion for BiTeI along the $K-\Gamma-H-A-L$ directions of the Brillouin zone. 
	The different electronic continua that are spectroscopically observed are indicated using the red and violet arrows. 
	}
\label{basic}%
\end{figure}

BiTeI is a polar, noncentrosymmetric semiconductor (space group \textit{P3m1}, point group $C_{3v}$) regarded as the archetypal quasi-2D Rashba system~\cite{PhysRevB.84.041202,Murakawa1490,maass2016spin}. 
It is a van der Waals-bonded layered structure, composed of a series of triangular network layers of I, Bi, and Te atoms stacked along its polar axis~\cite{article55,Tomokiyo_1977} [see Fig.\,\ref{basic}(b)]. 
Because of its layered structure, the bulk electronic bands are weakly dispersive along $k_{z}$~\cite{sakano2012threedimensional}. 
However, the band gaps and in-plane electronic dispersions within the $M-\Gamma-K$ ($k_{z}=0$) and $L-A-H$ ($k_{z}=\pi/c$) planes of the Brillouin zone (BZ) significantly differ from one another [see Fig.\,\ref{basic}(c)]. 
The principal conduction band (CB0), the highest valence band (VB0), and the next-lowest conduction band (CB1) all exhibit large Rashba splitting near the $\Gamma$ point and $A$ point of the BZ~\cite{PhysRevB.84.041202,PhysRevLett.108.246802} [see Fig.\,\ref{basic}(d)].
Despite being classified as a semiconductor, BiTeI is naturally doped due to nonstoichiometry~\cite{Tomokiyo_1977}; the chemical potential crosses CB0 and the charge carriers are electrons. 
Thus, BiTeI contains multiple Rashba continua due to the CB0\,$\rightarrow$\,CB0 and VB0\,$\rightarrow$\,CB0 transitions. 

BiTeI belongs to a materials class with a single polar axis of rotation \textbf{z} and a Rashba-like dispersion of the electronic bands
\begin{eqnarray}
E_{\eta}(k) = \frac{k^2}{2m^{*}} + \eta\alpha_{R}k,
\end{eqnarray}
where $m^{*}$ is the effective mass of the electron (assuming an isotropic, in-plane dispersion) and $k$ is the quasimomentum of the electrons within the normal plane of the polar axis $\textbf{xy}$, $\eta$ denotes the helicity of the subband, and the factor $\alpha_R = 2 E_{R} / k_{R}$ is the ``Rashba parameter'' that describes the magnitude of the $k$-dependent energy splitting [see Fig.\,\ref{basic}(a)]. 
$\eta = \pm$1 specifies whether the spin of the electron is parallel ($\eta = -1$) or antiparallel ($\eta = 1$) to $\textbf{k} \times \hat{z}$. 
The strength of $\alpha_{R}$ depends on the interaction of the total angular momentum of the electron with the electrical potential gradients close to heavy atomic nuclei within the lattice~\cite{PETERSEN200049,PhysRevLett.107.156803,PhysRevB.85.195401}. 

In this Letter, we use polarization-resolved resonant electronic Raman spectroscopy to study the continua of chiral excitations arising from transitions between helical electronic states in BiTeI. 
By scanning the energy of the incident photons, we selectively enhance distinct continua excitations at different points within the BZ [see Fig.\,\ref{basic}(c)]. 
The chiral excitations of the continua transform as the $A_2$ representation of the $C_{3v}$ lattice symmetry point group, which can only be observed in Raman spectra under resonant conditions, thus the dispersion of spin-polarized bands in a complex solid can be selectively studied by systematic scanning of the incident photon energy. 
By comparing the resonant Raman excitation profile of the continua to calculations within the three-band model, we deduce the Rashba parameters for the higher conduction bands. 
We also relate the spectroscopic data to {\it ab initio} density functional theory (DFT) calculations.

Single crystals of BiTeI were grown using the vertical Bridgman technique~\cite{C4CE01006J,kanou2013crystal}. 
Five-probe Hall effect and resistivity measurements were performed using a physical property measurement system (PPMS) by Quantum Design to determine the carrier concentration of the crystals, which was found to be $2.3\times 10^{19}$\,cm$^{-3}$. 

\begin{table}[b]
	\caption{\label{table:irreps}The Raman selection rules for BiTeI.}
	\setlength{\tabcolsep}{8pt}
	\setlength{\extrarowheight}{3pt}
	\vspace{5pt}
	\begin{tabular}{l l}
		\hline \hline
		Scattering Polarization & Irreducible\\ 
		Geometry & Representations\\
		\hline
		XX or YY & A$_1$ + E\\ 
		YX or XY & A$_2$ + E\\ 
		RR or LL & A$_1$ + A$_2$\\ 
		RL or RL & 2E\\ 
		\hline \hline
	\end{tabular} 
\end{table}

For polarization-resolved Raman scattering measurements we used a Kr$^+$ laser with photon energies ranging between 1.55 and 2.65\,eV in the quasi backscattering geometry. 
All spectra were acquired with 150 grooves/mm diffraction gratings (spectral resolution of $\sim$3\,meV) in a triple-stage spectrometer setup with a charge coupled device (CCD) detector. 
Single crystals of BiTeI were cleaved in a nitrogen-rich environment and then transferred into a continuous He-flow optical cryostat where all spectra were taken at 15\,K. 
The measured spectral intensities were corrected for the spectral response of the optical setup as well as the optical power absorption of the material medium. 

The scattering polarization geometries, defined by the incident and collected light polarizations, are labeled as $e_{i}e_{s}$, where $e_{i(s)}$ refers to the polarization of the incident (collected) light, respectively. 
The energies of the incident and collected photons are denoted as $\omega_i$ and $\omega_s$, respectively. 
The two optically distinct spatial directions that result from the layered structure lie either in the $\textbf{xy}$\,plane or along the $\textbf{z}$\,axis direction. 
The incident and collected photons were polarized in the $\textbf{xy}$\,plane and propagated along $\textbf{z}$. 
The secondary emission spectra contain both the photoluminescence (PL) and the Raman excitations. 
All of the measurements were performed in four scattering polarization geometries: XX, YX, RR, and RL. 
For linearly polarized light, the direction of the incident photon polarization is denoted as X, and the direction orthogonal to X in the $\textbf{xy}$\,plane is denoted as Y. 
R and L denote right and left circularly polarized light, respectively, such that $R(L) = X \pm iY$. 

In Table~\ref{table:irreps}, the Raman selection rules are listed for all four used scattering polarization geometries~\cite{koster1963properties}. 
The Raman-active excitations for the material with the $C_{3v}$ point group can be classified by the $A_{1}$, $A_{2}$, and $E$ irreducible representations. 
$A_{2}$ excitations are chiral because they are anti-symmetric with respect to the three vertical mirror reflections of the $C_{3v}$ point group and are the focus of this Letter. 
The resonant Raman scattering process can uniquely probe electronic excitations with an antisymmetric Raman tensor~\cite{ovander1960form,ovander1964form,shastry1990theory,SeanBook2022}. 

\begin{figure}[t]%
	\centering 
	\includegraphics[width=\linewidth]{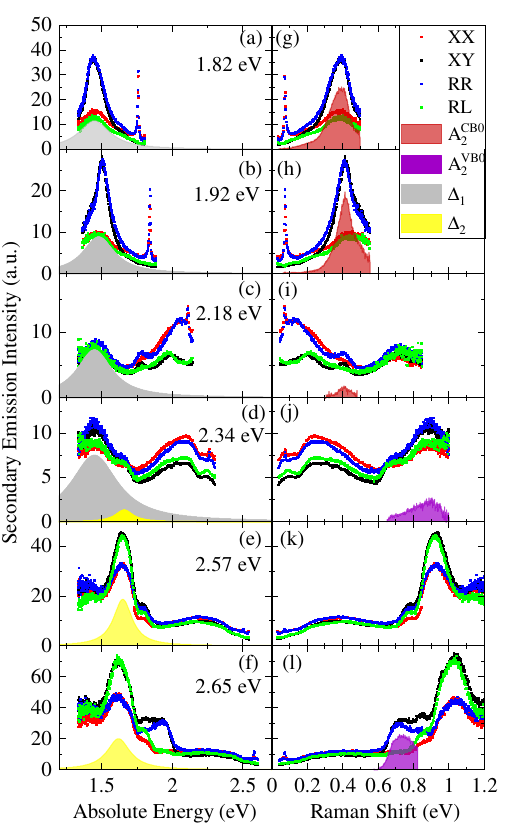}
	\caption{\textbf{Secondary emission spectra plotted against absolute energy (a)--(f) and Raman shift (g)--(l) for} $\boldsymbol{\omega}_{i}$ \textbf{between 1.83 and 2.65\,eV.} 
	The measured secondary emission spectra were acquired with XX (red), YX (black), RR (blue), and RL (green) scattering polarization configurations. 
	The PL features originating from the CB1\,$\rightarrow$\,CB0 and CB2\,$\rightarrow$\,CB0 transitions at the A-point are indicated by the shaded in regions of panels (a)--(f). 
	The $A_{2}$-symmetry Raman features due to CB0\,$\rightarrow$\,CB0 transitions near the $A$ point (red) and VB0\,$\rightarrow$\,CB0 transitions (violet) are indicated by the shaded in regions in panels (g)--(l). 
	The scales of $I(\omega_s)$ for each panel are chosen for the best readability. 
}
\label{SecEmm}%
\end{figure}

In Fig.\,\ref{SecEmm}, we plot the results of secondary emission measurements performed for the four scattering polarization geometries. 
PL features correspond to radiative electronic transitions at the $A$ point between CB1/CB2 and CB0, and their emission energy is independent of the excitation energy $\omega_{i}$. 
The energies of Raman-related features emitted at the scattering energy $\omega_{s}$ are fixed at the Raman shift $\omega_{R}=\omega_i-\omega_s$ for any $\omega_{i}$. 
To aid the reader's eye, the measured secondary emission cross section is plotted against $\omega_{s}$ for different $\omega_{i}$ in the panels on the left-hand side of Fig.\,\ref{SecEmm}, and plotted against the Raman shift on the right-hand side. 

The relevant spectral features are shaded in Fig.\,\ref{SecEmm}, with PL features on the left-hand side [Figs.\,\ref{SecEmm}(a)--\,\ref{SecEmm}(f)] and Raman related features on the right-hand side [Figs.\,\ref{SecEmm}(g)--\,\ref{SecEmm}(l)]. 
The shaded PL features are $\Delta_{1}=1.48$\,eV and $\Delta_{2}=1.65$\,eV. 
The distinct Raman features with $A_2$ symmetry are labeled as $A_{2}^{B}$, where $B$ denotes the excited band~\footnote{As the energy of the $A_2$ excitations are more than an order of magnitude higher than any of the phonon bands~\cite{sklyadneva2012lattice,gnezdilov2014enhanced}, we identify the transitions as of electronic origin.}. 

\begin{table}[t]
	\caption{\label{table:PL}The lowest energy PL features in  BiTeI.}	
	\setlength{\tabcolsep}{8pt}
	\setlength{\extrarowheight}{3pt}
	\vspace{5pt}
	\begin{tabular}{l l l l}
		\hline \hline
		PL & Energy  & Energy  & Transition \\ 
		Feature & (meas.) & (DFT) & (A-point)\\
		\hline
		$\Delta_{1}$ & 1.48\,eV & 1.52\,eV & CB1$\rightarrow$CB0\\ 
		$\Delta_{2}$ & 1.65\,eV & 2.05\,eV & CB2$\rightarrow$CB0\\ 
		\hline \hline
	\end{tabular} 
\end{table}

In Table~\ref{table:PL}, we list the PL features, the electronic transitions they correspond to, and compare the measured and DFT calculated energy differences from the CB0 band at the \textit{A}-point [Fig.\,\ref{basic}(d)]. 
The CB1\,$\rightarrow$CB0\,transition occurs at the $A$ point and not between the true CB1 minima and the lower CB0 subband because the CB0 states involved in such a transition are occupied; transitions to the higher CB0 sub-band fall outside the spectral range of our experimental setup. 
The CB2\,$\rightarrow$\,CB0 transition occurs at the $A$ point, where CB2 is the next-next lowest conduction band. 

The $\Delta_{2}$ spectral intensity undergoes a resonant enhancement near $\omega_{i}=2.57$\,eV. 
To account for this resonance, we note that the VB1\,$\rightarrow$\,CB2 energy gap, where VB1 is the next-highest valence band, is overestimated by DFT. 
The DFT-calculated VB1\,$\rightarrow$\,CB2 energy gap is 1.3\,eV, however, the angle-resolved photoemission spectroscopy (ARPES) measurement places the VB1\,$\rightarrow$\,CB0 gap at about $\sim$1.0\,eV~\cite{sakano2012threedimensional}. 
The gap measured by emission $\Delta_{2}$ is 1.65\,eV compared to its calculated value at 2.0\,eV (see Table~\ref{table:PL}). 
Therefore, the true VB1\,$\rightarrow$\,CB2 energy gap is about $\sim$2.65\,eV, and the resonance near $\omega_{i}=2.57$\,eV is due to the VB1\,$\rightarrow$\,CB2 transition at the $A$ point. 
In addition, the $\Delta_{2}$ feature appears partially polarized under resonance, favoring light emission in YX and RL over XX and RR polarizations. 
The transition selection rules for the states near the $A$ point cannot account for its polarized nature. 
This implies that VB1 and CB2 bands are spin polarized away from the $A$ point. 

\begin{figure}[t]%
	\centering
	\includegraphics[width=1.0\linewidth]{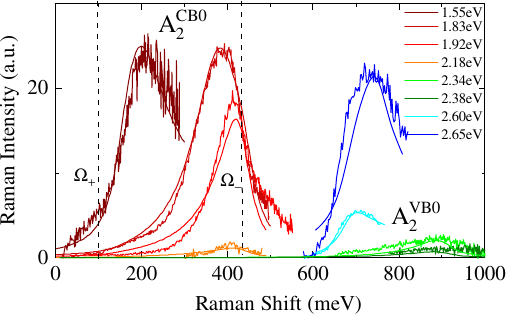}
	\caption{\textbf{The $A_{2}$ Raman excitations overlaid with calculations of the $A_2$ Raman response.} 
		The $A_{2}$ Raman excitations for $\omega_i = 1.55--2.65$\,eV are labeled as $A_{2}^{CB0}$ and $A_{2}^{VB0}$ to denote which Rashba continua they are originate from. 
		The $A_{2}^{CB0}$ Raman excitations are colored red and the $A_{2}^{VB0}$ Raman excitations are colored green and blue. 
		The calculated bounds of the $A_{2}^{CB0}$ Rashba continuum $\Omega_{\pm}$ are indicated by the dotted lines. 	
		The $A_{2}$ Raman response is calculated for $\omega_i$ that coincides with the laser photon energy. 
		A decay width of $\gamma=15$\,meV is used for the energy states. 
	}
	\label{excit}%
\end{figure}

We now turn to the Raman excitations with $A_{2}$ symmetry [see Figs.\,\ref{SecEmm}(g)--\,\ref{SecEmm}(l)]. 
The $A_{2}$ Raman excitations can be deduced from the scattering polarization geometries by noting that $A_{2}$ excitations are present with equal intensity in XY and RR but are absent in XX and RL scattering polarizations (see Table~\ref{table:irreps}). 
The $A_{2}$ Raman excitation features are plotted in Fig.\,\ref{excit} and are labeled as either $A_{2}^{CB0}$ or $A_{2}^{VB0}$. 
All of the $A_{2}$ Raman excitations at frequencies $\omega_{R}$ shift with increasing incident photon energy $\omega_{i}$. 
The $A_{2}^{CB0}$ and $A_{2}^{VB0}$ electron-hole excitations fall within the limits of the Rashba continua excited under resonant conditions involving both CB1 and CB2 bands as the intermediate states (see Table~\ref{table:chiral}).  

Since the wave functions of the $A_{2}$ excitations are antisymmetric with respect to the exchange of the $\textbf{x}$ and $\textbf{y}$ axes, there are two (left- and right-handed) enantiomeric isomers of the $A_{2}$ excitation. 
However, the charge density distributions corresponding to these excitations are identical under the mirror reflections of $C_{3v}$. 
Thus, the two isomers are indistinguishable from one another in RR or LL scattering spectra.

\begin{table}[b]
	\caption{\label{table:chiral} The $A_{2}$ Raman continua of BiTeI.}
	\setlength{\tabcolsep}{8pt}
	\setlength{\extrarowheight}{3pt}
	\vspace{5pt}
	\begin{tabular}{l l l}
		\hline \hline
		A$_{2}$ Continuum & A$_{2}^{CB0}$ & A$_{2}^{VB0}$\\ 
		\hline
		Energy Range & 0.1-0.4\,eV & $>$0.4\,eV\\
		Transition & CB0$\rightarrow$CB0 & VB0$\rightarrow$CB0\\
		Intermediate Bands & CB1/CB2 & CB1/CB2 \\
		\hline \hline
	\end{tabular} 
\end{table}

To verify the assignment of the $A_2$ signal to the continua of electronic transitions between Rashba split bands, we calculated the high-energy Raman scattering spectra throughout the BZ. 
The general expression for the differential light scattering cross section per unit scattered frequency per unit solid angle is $d^2\sigma/d\omega_{s}d\Omega  \propto \frac{\omega_{s}}{\omega_{i}} R$ \cite{RevModPhys.79.175}, where $R$\,is the transition rate given by Fermi's golden rule,
\begin{eqnarray}
R \propto \int d^{2}\textbf{k} \frac{\omega_{s}}{\omega_{i}} |H_{f,i}|^{2} \delta(\Tilde{\epsilon}_{f,\textbf{k}} - \Tilde{\epsilon}_{i,\textbf{k}} - \omega_{R}) ,
\end{eqnarray}
where $\Tilde{\epsilon}_{s,\textbf{k}} = \epsilon_{s}(\textbf{k}) + i\gamma$ are the energy eigenvalues for the initial ($i$), intermediate ($m$), and final ($f$) states, and $\gamma$ is the decay width. 
We retain only the resonant part of the interaction Hamiltonian $H_{f,i}$. 
Non-resonant terms do not contribute to the cross section for A$_2$ symmetry excitations~\cite{ovander1960form,ovander1964form}. 
$H_{f,i}$ connects the initial $\ket{i}$ and final $\ket{f}$ states of the system via intermediate electronic states $\ket{m}$~\cite{cardona1983lightscattering}, 
\begin{eqnarray}
H_{f,i} \approx \sum_{m\neq i} \frac{\bra{f}\hat{H}_{int}\ket{m}\bra{m}\hat{H}_{int}\ket{i}}{\Tilde{\epsilon}_{m,\textbf{k}} - \Tilde{\epsilon}_{i,\textbf{k}} - \omega_i} 
\end{eqnarray}

The energy eigenvalues of the electronic states used in these calculations are determined as follows:  
The energy eigenvalues of the initial (final) states $\Tilde{\epsilon}_{i(f),\textbf{k}}$ are estimated using the dispersions of VB0 and CB0 measured by ARPES~\cite{ishizaka2011giant,sakano2012threedimensional}. 
The intermediate-state energy eigenvalues $\Tilde{\epsilon}_{m,\textbf{k}}$ are chosen to fit the $A_{2}$ Raman data, where CB1 has a Rashba-like dispersion and CB2 has a parabolic dispersion near the $A$ point. 
The in-plane dispersion near the $A$ point is practically isotropic~\cite{ishizaka2011giant}. 
We approximate the dispersion within the $K-\Gamma-M-K$ area by interpolating between the $\Gamma-K$ and $\Gamma-M$ dispersions as $\hat{k}$ changes from $\Gamma-K$ to $\Gamma-M$ using a periodic function to reproduce the hexagonal in-plane periodicity. 
Finally, we connect the two dispersions by a cosine function that reproduces the periodicity in the $k_{z}$ direction. 

\begin{table}[b]
	\caption{\label{table:parameters} Higher conduction band parameters of BiTeI near the $A$ point of the BZ using the three-band model calculation, and corresponding parameters given by DFT. The in plane effective mass m$^{*}$ is represented as a fraction of the rest mass of the electron $m_{e}$.}
	\setlength{\tabcolsep}{8pt}
	\setlength{\extrarowheight}{3pt}
	\vspace{5pt}
	\begin{tabular}{l l l}
		\hline \hline
		Parameter & CB1 & CB2\\
		 \hline
		$\alpha_{R}$ (meas.) & 1.35\,eV \AA & $\sim$0~\footnote{If the CB2 band is Rashba spin-split, then its value of $\alpha_{R}$ is below the analysis uncertainty.}\\  
		$\alpha_{R}$ (DFT) & 1.15\,eV \AA & 0\\  
		m$^{*}$ (meas.) & 0.15m$_{e}$ & 0.14m$_{e}$\\  
		m$^{*}$ (DFT) & 0.20m$_{e}$ & 0.34m$_{e}$\\  
		\hline \hline
	\end{tabular} 
\end{table}

The position of the Fermi energy, $E_{F}$, determines the bounds of the Rashba continuum~\cite{PhysRevB.91.035106} because the points where $E_{F}$ intersects CB0 determine the lowest ($\Omega_{+}$) and the highest ($\Omega_{-}$) transitions of the CB0\,$\rightarrow$\,CB0 continuum [see Fig.\,\ref{basic}(a)]. 
We deduce the position of $E_{F}$ by relating it to the carrier concentration determined from Hall effect measurements~\cite{lee2011optical}. 

In Fig.\,\ref{excit}, we display model calculations overlaid with $A_{2}$ Raman data for a set of incoming excitation photon energies $\omega_{i}$ between 1.55 and 2.65\,eV. 
The calculation captures the slightly asymmetric character of the otherwise Lorentzian-like line shape, which is due to the resonant denominator $(\epsilon_{m,\textbf{k}} - \epsilon_{i,\textbf{k}} - \omega_i)^{2} + \gamma^{2}$. 
More importantly, the calculation of the $A_{2}^{CB0}$ continuum captures the large shift of the Raman peak signal with increasing $\omega_{i}$, which is due to the joint density of states of transitions between CB0 subbands. 
The $A_{2}^{VB0}$ continuum has a single resonance at the principle band gap between VB0 and CB0, which occurs around $k_{xy}=k_{R}^{VB0}$. 
Although the $A_{2}^{VB0}$ continuum can, in principal, begin at $\omega_{R}=0.3$\,eV, Pauli blocking due to occupied CB0 states pushes the lower limit of the continuum to higher energies, depending on the carrier concentration of the sample. 
The strong $A_{2}^{VB0}$ feature for $\omega_{i}=$2.60 and 2.65\,eV excitations are due to transitions that occur near the $k_{z}=\pi/c$ plane of the BZ. 
Thus, the calculation provides good confirmation that the observed spectral features correspond to Raman-active intersubband transitions between Rashba-split bands. 

Finally, we compare DFT calculations of CB1 and CB2 to the dispersion acquired through the Raman scattering calculation (see Table~\ref{table:parameters}). 
Near the $\Gamma$ point, the CB1 and CB2 dispersions are close to their values calculated by DFT. 
In the proximity of the $A$ point, the CB1 Rashba dispersion parameters, including $\Delta_{1}$, $\alpha_{R}$, and $m^{*}_{CB1}$, are also close to the values calculated by DFT. 
However, the CB2 parabolic dispersion parameters differ from the DFT calculated ones; $\Delta_{2}$ and m$^{*}_{CB2}$ are 80\% and 41\% of their respective values calculated by DFT. 
The disparity between deduced CB2 parameters and the dispersion given by DFT is surprising considering that the CB1 and CB0 parameters are well described by DFT. 
This is further evidence that some aspect of the CB2 band is not accounted for in the DFT calculation. 

In conclusion, we probed the continua of electronic transitions between Rashba-split helical bands in the quasi-2D Rashba system BiTeI. 
The continua belong to the antisymmetric $A_{2}$ irreducible representation of the $C_{3v}$ point group, thus the excitations are chiral. 
The quasi-2D nature of the electronic structure allows for the selective excitation of chiral electronic continua originating from different areas of the BZ, which is only possible by exploiting the resonant Raman scattering effect. 

By comparing spectroscopic data to calculations within a three-band model, the CB1 band was confirmed to have a Rashba dispersion with parameters similar to those given by DFT: $\Delta_{1} = 1.48$\,eV, $\alpha_{R}=1.35$\,eV \AA\,,and $m^{*}=0.15m_{e}$. 
The CB2 parabolic dispersion differs significantly from DFT calculations: $(\Delta_{2}$, m$^{*})=(1.65\,$eV$, 0.14m_{e})$ vs. $(2.05\,$eV$, 0.34m_{e})$ as calculated by DFT. 
Under resonance, the CB2\,$\rightarrow$\,CB0 transition appears partially polarized, suggesting that CB2 is also spin polarized away from the $A$ point. 

More generally, the presented approach of studying BiTeI may enable control of chiral electronic excitations in systems with a strong Rashba-like dispersion and offers critical insight into novel chirality-based phenomena in quasi-2D Rashba systems. 

We would like to thank Z. Lin for enlightening and useful discussions. 
The spectroscopic work at Rutgers (A.C.L., H.-H.K. and G.B.) was supported by the NSF under Grants No. DMR-1709161 and No. DMR-2105001. 
The characterization work (K.D. and S.-W.C) was supported by the DOE under Grant No. DOE: DE-FG02-07ER46382. 
The crystal growth at Postech was supported by The National Research Foundation of Korea (NRF) funded by the Ministry of Science and ICT, grant No. 2020M3H4A2084417. 
B.M. acknowledges support from the Gianna Angelopoulos Programme for Science, Technology, and Innovation and from the Winton Programme for the Physics of Sustainability.
B.P. acknowledges support from the Winton Programme for the Physics of Sustainability.
The calculations were performed using resources provided by the Cambridge Tier-2 system, operated by the University of Cambridge Research Computing Service and funded by EPSRC Tier-2 Capital Grant No. EP/P020259/1, as well as with computational support from the U.K. Materials and Molecular Modelling Hub, which is partially funded by EPSRC (EP/P020194), for which access is obtained via the UKCP consortium and funded by EPSRC Grant ref. EP/P022561/1. 
The work at NICPB (G.B.) was supported by the European Research Council (ERC) under the European Union{\textquoteright}s Horizon 2020 research and innovation programme Grant Agreement No. 885413. 

%
%
%

%
\end{document}